\documentclass{article}

\usepackage{PRIMEarxiv}

\usepackage[utf8]{inputenc} 
\usepackage[T1]{fontenc}    
\usepackage{hyperref}       
\usepackage{url}            
\usepackage{booktabs}       
\usepackage{amsfonts}       
\usepackage{nicefrac}       
\usepackage{microtype}      
\usepackage{lipsum}
\usepackage{fancyhdr}       
\usepackage{graphicx}       
\graphicspath{{media/}}     
\usepackage{xcolor}
\usepackage{amsmath}
\usepackage{amsfonts}

\usepackage{amsmath,tabu,booktabs}
\usepackage{stfloats}
\title{A graph-space optimal transport FWI approach \\ based on $\kappa$-generalized Gaussian distribution
\thanks{\textit{\underline{Citation}}: 
\textbf{da Silva, S.L.E.F. \& Kaniadakis, G. (2023) A graph-space optimal transport FWI approach based on $\kappa$-generalized Gaussian distribution, in Third International Meeting for Applied Geoscience \& Energy Expanded Abstracts. December 2023, 670-674, Houston, TX, USA. DOI: 10.1190/image2023-3916733.1.}} 
}

\author{
  Sérgio Luiz E. F. da Silva$^{1,2,3}$ \& G. Kaniadakis$^{1,2,}$ \\
  $^{1}$Dipartimento di Scienza Applicata e Tecnologia (DISAT), Politecnico di Torino, Turin, Italy\\
  $^{2}$Institute for Complex Systems of the National Research Council c/o  Politecnico di Torino, Turin, TO, Italy \\
  $^{3}$GISIS, Fluminense Federal University, Niter\'oi, RJ, Brazil \\
  \texttt{sergio.dasilva@polito.it; giorgio.kaniadakis@polito.it} 
}

\begin{document}
\maketitle

\begin{abstract}
The statistical basis for conventional full-waveform inversion (FWI) approaches is commonly associated with Gaussian statistics. However, errors are rarely Gaussian in non-linear problems like FWI. In this work, we investigate the portability of a new objective function for FWI applications based on the graph-space optimal transport and $\kappa$-generalized Gaussian probability distribution. In particular, we demonstrate that the proposed objective function is robust in mitigating two critical problems in FWI, which are associated with cycle skipping issues and non-Gaussian errors. The results reveal that our proposal can mitigate the negative influence of cycle-skipping ambiguity and non-Gaussian noises and reduce the computational runtime for computing the transport plan associated with the optimal transport theory.
\end{abstract}


\section{Introduction}
Full-Waveform Inversion (FWI) is a non-linear inverse problem that aims to obtain subsurface physical parameters (e.g., sound speed, attenuation coefficient, or density) by comparing modeled and observed data. The modeled and observed data are compared using an objective function, usually based on the least-squares method \cite{VirieuxOperto2009}. FWI is a robust methodology that significantly outperforms standard geophysical-data inversion methods because it employs all the information in seismic waveforms (phase, amplitude, and complete shape of waveforms) rather than simply wave amplitudes or travel-time arrivals. Indeed, several cases of successful FWI applications on an industrial scale have already been reported in the literature.

Although FWI is a powerful approach, some questions remain as open problems in which several issues are often associated with the objective function. Two significant challenges are cycle-skipping and seismic noise's non-Gaussianity. Mitigating the cycle-skipping effects (or phase ambiguity) has recently been hotly debated. Consequently, several convex objective functions concerning travel-time misfits have been proposed, such as (de)convolution-based objective functions \cite{deconvolution_luo_sava_2011}, extended-domain FWI frameworks \cite{WRI_Tristan_2013}, machine learned functions \cite{doi:10.1190/segam2020-3427351.1}, as well as optimal transport approaches \cite{Metivier_GraphSpace_OT_2018,daSilvaetal2022GJI,daSilva2023Entropy}; the latter being one of the approaches considered in this work.

In the 18th century, G. Monge raised the transportation theory to study the optimal allocation of resources. Since then, several formulations of the optimal transport (OT) problem have been developed over the last decades. In this work, we consider a well-posed relaxed version of the OT formulation introduced by L. Kantorovich. In this approach, the metric associated with the OT is based on the Wasserstein distance, also known as the Kantorovich-Rubinstein metric. However, such a metric is valid for comparing probability distributions;  this is not the case for seismic signals that are oscillatory and non-normalized. Thus, we consider the waveforms in the graph space to satisfy the axioms of probability theory.

Numerous proposed OT approaches in the literature have considered Gaussian statistics as pillars of their theories. For example, the most popular approach to FWI is based on the $L_2$-norm (hereafter, conventional FWI), which assumes that the difference between modeled and observed data (errors) follows a Gaussian distribution. Although the conventional FWI approach has been widely used, the assumption that errors follow Gaussian statistics is not always appropriate \cite{crase1990}, especially in nonlinear problems \cite{tarantola2005book} like FWI. For this reason, different objective functions based on non-Gaussian criteria have been proposed in the literature \cite{tristanTstudentSEG2011,carvalho_et_al_2022_GJI_229}. In this regard, non-Gaussian distributions based on the so-called $\kappa$-generalized Gaussian distribution is a robust alternative. The $\kappa$-Gaussian distribution has been considered in several contexts such as in economics \cite{e15093471} and inverse problems \cite{10.1371/journal.pone.0282578}.

In this work, we propose a robust FWI formulation based on a $\kappa$-generalized Gaussian distribution (or $\kappa$-Gaussian distribution) and the OT distance in the Kantorovich-
Rubinstein sense. Starting from the principle of maximum likelihood, we construct a new outlier-resistant objective function and place it in the context of the optimal transport theory. In this way, our proposal embraces robust properties against non-Gaussian noise while mitigating cycle-skipping effects.

\section{Theory}

\subsection{Conventional FWI formulation}

FWI retrieves a subsurface model $m$ by matching modeled data with observed data, usually employing a least-squares method \cite{VirieuxOperto2009}. The least-squares FWI (conventional FWI) can be formulated, in the time-domain, as the following optimization problem:
\begin{equation}
    \underset{m}{min} \,\, \phi_{0} (m) :=
    \frac{1}{2} \sum_{s,r} \int_{0}^{T}
    \Big(\Gamma_{s,r} u_s(m,t) - d_{s,r}(t)\Big)^2 dt,
\label{eq:FWI_classical_objective_function}
\end{equation}
where $\phi_0$ represents the least-squares objective function, $t \in [0,\,T]$ denotes the temporal variable in which $T$ is the acquisition time, $\Gamma_{s,r} u_s$ and $d_{s,r}$ are, respectively, the modeled and observed data with $\Gamma_{s,r}$ representing a sampling operator that extract modeled waveforms associated with the receiver $r$ and seismic source $s$. 

In this work, we consider the propagation of acoustic waves; therefore, the seismic wavefields are the pressure fields $u_s$ that satisfy the acoustic wave equation,
\begin{equation}
\nabla^2 u_s(\textbf{x},t) 
- m (\textbf{x}) \frac{\partial^2 u_s (\textbf{x},t)}{\partial  t^2 }   
= g(t) \delta(\textbf{x}-\textbf{x}_s),
\label{eq:waveeqtime}
\end{equation}
where $\nabla^2$ is the Laplace operator, $\textbf{x}$ represents spatial coordinates, and $g(t) \delta(\textbf{x}-\textbf{x}_s)$ denotes the seismic source $s$ at the position $\textbf{x} = \textbf{x}_s$, in which $\delta$ is the Dirac Delta function, and the model parameters are the squared slowness: $m =\frac{1}{c^2}$, with $c$ representing the P-wave velocity.

Let us define the error as the difference between modeled and observed data, $\varepsilon = \Gamma_{s,r} u_s - d_{s,r}$. The conventional FWI assumes that each error $\varepsilon_{i}$ is an independent normal random variable. Thus, minimizing the least-squares objective function \eqref{eq:FWI_classical_objective_function} is equivalent to obtaining the maximum Gaussian likelihood following \cite{tarantola2005book}:
\begin{equation}
     \underset{m}{min} \,\, \phi_{0} (m) \equiv \underset{m}{max} \,\, 
 \mathcal{L}_G (m) ,
    \label{eq:gaussian_likelihood}
\end{equation}
where the Gaussian likelihood can be defined as 
\begin{equation}
    \mathcal{L}_G(m) \propto  
     \prod_{s,r}  \exp \Bigg(-\frac{1}{2} \int_{0}^{T}
    \varepsilon_{s,r}^2(m,t) dt \Bigg).
    \label{eq:gaussian_likelihood}
\end{equation}
Therefore, it is remarkable that determining a model $m$ that minimizes the least-squares objective function \eqref{eq:FWI_classical_objective_function} is equivalent to obtaining a model that maximizes the Gaussian likelihood \eqref{eq:gaussian_likelihood} or minimizes the negative log-likelihood:
\begin{equation}
     \underset{m}{min} \,\, \phi_{0} (m) \equiv \underset{m}{max} \,\, 
 \mathcal{L}_G (m) \equiv \underset{m}{min} \,\, -ln\Big[ 
 \mathcal{L}_G (m)\Big],
    \label{eq:gaussian_likelihood_log}
\end{equation}
where $\ln(\cdot)$ is the natural logarithm. We invite the reader to consult section 2 of Ref. \cite{DASILVA2022127554} for more mathematical details.

\newpage

\subsection{FWI based on $\kappa$-Gaussian distribution}

Because the statistical distribution of the errors in non-linear issues like FWI is typically non-Gaussian \cite{crase1990,tarantola2005book}, we assume that the errors are independent and randomly distributed by a non-Gaussian probability function known as the $\kappa$-generalized Gaussian distribution \cite{PhysRevE.101.053311}. Such probability distribution arises from the maximization of so-called $\kappa$-entropy \cite{KANIADAKIS2001405,PRE2002,PRE2005}, which has been widely used in a wide variety of scientific and technological problems \cite{entropy2024}. This probability function is a one-parameter generalization of the Gaussian distribution that can be defined as:
\begin{equation}
    p_\kappa(\varepsilon_i) = \frac{1}{Z_\kappa} \exp_\kappa\left(-\beta_\kappa\,\varepsilon_i^2\right),
    \label{eq:kappa_Gaussian_general_form}
\end{equation}
where the normalization constant $Z_\kappa$ and the $ \beta_\kappa$-factor are given by \cite{DASILVA2022127554}:
\begin{equation}
    Z_\kappa = \sqrt{\frac{\pi}{\beta_\kappa}} \frac{|2\kappa|^{-1/2}}{1+\frac{1}{2}|\kappa|} \frac{\Gamma\left(\frac{1}{|2\kappa|} - \frac{1}{4}\right)}{\Gamma\left(\frac{1}{|2\kappa|} + \frac{1}{4}\right)}
    \label{eq:normalization_constant_2}
\end{equation}
and
\begin{equation}
    \beta_\kappa = \frac{|2\kappa|^{-1}}{2} \frac{1+\frac{1}{2}|\kappa|}{1+\frac{3}{2}|\kappa|} \frac{\Gamma\left(\frac{1}{|2\kappa|} - \frac{3}{4}\right) \, \Gamma\left(\frac{1}{|2\kappa|} + \frac{1}{4}\right)}{\Gamma\left(\frac{1}{|2\kappa|} + \frac{3}{4}\right) \, \Gamma\left(\frac{1}{|2\kappa|} - \frac{1}{4}\right)}
    \label{eq:beta_kappa}
\end{equation}
holding for $|\kappa| < 2/3$, while the $\kappa$-generalized exponential function is defined as follows \cite{KANIADAKIS2001405}:
\begin{equation}
    \exp_\kappa (y) = \Big(\sqrt{1+\kappa^2 y^2}+\kappa y \Big)^\frac{1}{\kappa}.
    \label{eq:kappaExponential}
\end{equation}

The Gaussian distribution, 
\begin{equation}
    p_0(\varepsilon_i) = \frac{1}{\sqrt{2\,\pi}} \exp\left(-\frac{1}{2}\,\varepsilon_i^2\right),
    \label{eq:Gaussian_general_form}
\end{equation}
is a particular case of the $\kappa$-Gaussian distribution \eqref{eq:kappa_Gaussian_general_form} in the limit $\kappa \rightarrow 0$, since in this classical limit $Z_0 = \sqrt{2\,\pi}$, $\beta_0 = 1/2$, and $\exp_0(y) = \exp(y)$ is the well-known exponential function.

\begin{figure*}[!b]
  \centering
  \flushleft{(a) \hspace{3.8cm} (b) \hspace{3.45cm} (c) \hspace{3.55cm} (d)}
  \includegraphics[width=\textwidth]{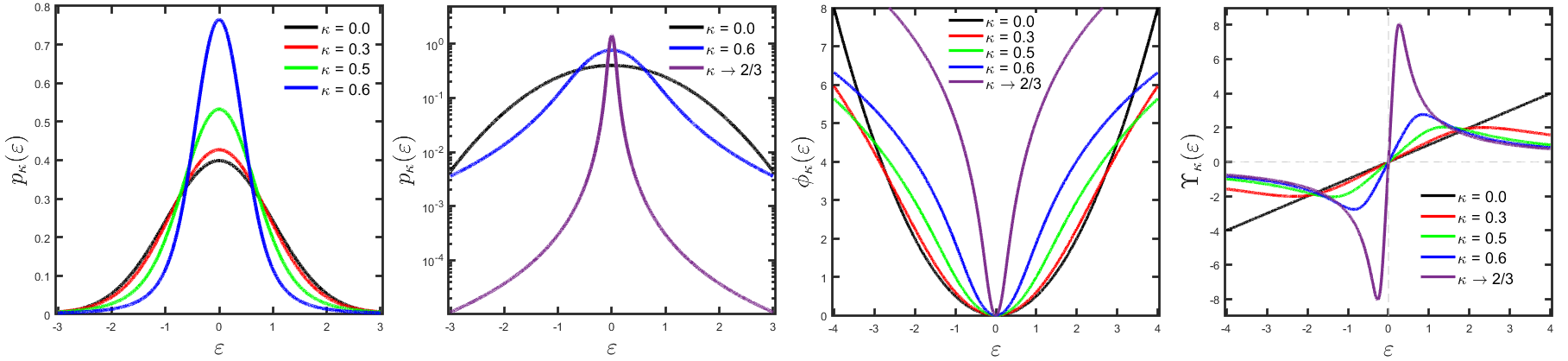}
  \caption{Probability plots of the $\kappa$-Gaussian distribution for some $\kappa$-values using (a) a linear scale and (b) a linear scale on the axis of abscissas, and a logarithmic scale on the axis of ordinates, in which the black line represents the standard Gaussian distribution ($\kappa \rightarrow 0$). (c) Graphical representation of the $\kappa$-objective function for some $\kappa$-values, and (d) the respective influence function, where the black line represents the classical approach ($\kappa \rightarrow 0$).}
    \label{fig:k_gaussian_objective}
\end{figure*}

Figures \ref{fig:k_gaussian_objective}(a) and \ref{fig:k_gaussian_objective}(b) depict the plots of the $\kappa$-Gaussian distribution \eqref{eq:kappa_Gaussian_general_form} for typical $\kappa$-values, with the solid black curve referring to the standard Gaussian distribution ($\kappa \rightarrow 0$), using (a) a linear scale and (b) a linear-logarithmic scale. Note that as the $\kappa$-value increases, which means a greater increase from a Gaussian behavior, the fat tails are observed. This means that the $\kappa$-Gaussian distribution considers a higher weight for smaller errors while applying a lower weight for more significant errors. 

Now, by assuming that the errors are independent and randomly distributed by the non-Gaussian distribution in Eq.~\eqref{eq:kappa_Gaussian_general_form}, we can obtain the $\kappa$-objective function $\phi_\kappa$ employing the maximum likelihood method:
\begin{equation}
    \min_m \phi_\kappa(m) \propto \max_m \mathcal{L}_\kappa (m) := \prod _{i = 1}^{N} p_\kappa\Big(\varepsilon_i (m)\Big),
\end{equation}
where $\mathcal{L}_\kappa$ represents the $\kappa$-likelihood function. Then, by minimizing the negative log-likelihood, we obtain the $\kappa$-objective function as follows:
\begin{equation}
    \phi_\kappa(m) = - \sum_{s,r} \int_0^T \ln\Bigg[\exp_\kappa\Big(-\beta_\kappa \varepsilon_{s,r}^2(m)\Big)\Bigg] dt.
    \label{eq:kappaCostFunction}
\end{equation}
We emphasize that in the limit $\kappa \rightarrow 0$, the $\kappa$-objective function converges to the conventional least-squares objective function $\phi_0$ \eqref{eq:FWI_classical_objective_function}. In Fig.~\ref{fig:k_gaussian_objective}(c) we illustrate the $\kappa$-objective function for some $\kappa$-values, where the black curve corresponds to the conventional case.

The $\kappa$-objective function \eqref{eq:kappaCostFunction} is based on a non-Gaussian criterion and, therefore, it is robust concerning erratic measurements (outliers).
In order to determine the $\kappa$-objective function's sensitivity to outliers, we compute the influence function, which is defined, for a given model $m_j$, as:
\begin{equation}
    \Upsilon_\kappa := \frac{\partial \phi_\kappa(\varepsilon; m_j)}{\partial \varepsilon}.    \label{eq:inflencefunctiondef}
\end{equation}
If $\Upsilon \rightarrow 0$ under $\varepsilon \rightarrow \pm\infty$, the related objective function is said to be robust, and non-robust if $\Upsilon \rightarrow \pm\infty$ under $\varepsilon \rightarrow \pm\infty$ \cite{BookInfluenceFunction_Hampel2005}. The influence function of the $\kappa$-objective function is given by:
\begin{equation}
    \Upsilon_\kappa =
    \begin{cases}
    \frac{2\beta_\kappa \varepsilon}{\sqrt{1 + \kappa^2 \beta_\kappa^2 \varepsilon^4}}, & \text{for} \quad  0 < \kappa < 2/3 \\
    \varepsilon, & \text{for} \quad  \kappa = 0.
  \end{cases}
    \label{eq:kappainflencefunction}
\end{equation}
According to Eq. \eqref{eq:kappainflencefunction}, the influence function tends to infinity in the limit $\kappa \rightarrow 0$ if there is an outlier ($\varepsilon \rightarrow \infty$) in the dataset (black curve in Fig.~\ref{fig:k_gaussian_objective}(d)). In contrast, the influence function tends to zero in the $0 < \kappa < 2/3$ case if there is an outlier in the dataset (colorful curves in Fig.~\ref{fig:k_gaussian_objective}(d)).

\subsection{A $\kappa$-graph-space optimal transport FWI approach}

Let us consider a seismic trace $d(t)$ as a set of ordered pairs $\{(t_i,d_i) \in \mathbb{R}^{2}, i = 1, 2, \cdots, N\}$ with $d_i = d(t_i)$, in which $t_i$ represents the time discretization. The $\kappa$-objective function based on the Wasserstein distance between the graph-transformed \cite{Metivier_GraphSpace_OT_2018} quantities $d_{mod,i}^{\mathcal{G}_\kappa} = (t_i,d^{mod}_i)$ and $d_{obs,i}^{\mathcal{G}_\kappa} = (t_i,d_i^{obs})$ reads \cite{PhysRevE.106.034113}:
\begin{equation}
    \underset{m}{min } \quad \phi_{\mathcal{W}_\kappa} (m) :=  \sum_{s,r}  \mathcal{W}_\kappa^{\mathcal{G}_\kappa}\Big(d_{s,r}^{mod}(m,t), d_{s,r}^{obs}(t)\Big),
    \label{eq:finalcostfunction}
\end{equation}
with
\begin{equation}   \mathcal{W}_\kappa^{\mathcal{G}_\kappa} \Big(d_{mod}^{\mathcal{G}_\kappa}, d_{obs}^{\mathcal{G}_\kappa} \Big) = \underset{\sigma \in \Omega(N)}{min } \quad -\sum_{i=1}^{N_t} \ln\Big(\Phi_\kappa(t_i,\sigma)\,\,\Psi(d_i,\sigma)\Big),
    \label{eq:k_GSOT_misfitfunction_01}
\end{equation}
where
\begin{equation}
\Phi_\kappa(t_i,\sigma) = \exp_\kappa\Bigg[-\beta_\kappa \Big(t_i - t_{\sigma(i)}\Big)^2\Bigg]
\end{equation}
and
\begin{equation}
    \Psi(d_i,\sigma) = \exp_\kappa\Bigg[-\beta_\kappa \Big(d_i^{mod} - d^{obs}_{\sigma(i)}\Big)^2\Bigg],
\end{equation}
in which $\sigma$ is obtained by solving a linear sum
assignment problem (LSAP) and $N_t$ is the time samples. We call our proposal by $\kappa$-Graph-Space Optimal Transport FWI ($\kappa$-GSOT-FWI).

\newpage
\section{Numerical experiments}

\begin{figure}[!htb]
  \centering
  \includegraphics[width=.5\columnwidth]{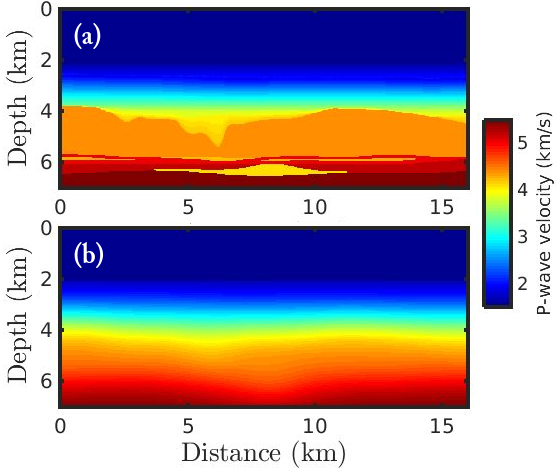}
  \caption{(a) Benchmark model, representing the deepwater Brazilian pre-salt region. (b) Initial model used in FWI tests.}
    \label{fig:usedmodels}
\end{figure}

To demonstrate the potentialities of the $\kappa$-GSOT-FWI in mitigating the negative impacts due to non-Gaussian noises and cycle-skipping issues, we perform several numerical simulations involving a time-domain FWI in a typical deepwater Brazilian pre-salt setting. The benchmark model, namely Chalda, is depicted in Fig. \ref{fig:usedmodels}(a). The seismic survey comprises 161 seismic sources equally spaced every 50 m, at 10 m in-depth, and are 7Hz Ricker wavelets. In addition, we simulate an ocean bottom node (OBN) acquisition by employing  21 receivers on the ocean floor equally spaced every 400 m. At a sample rate of 2 ms, we take 7 s into account for the seismic collection time. Additionally, we apply a high-pass filter on the seismic dataset to eliminate energy below 2.5 Hz in order to recreate a realistic case.

Figure \ref{fig:usedmodels}(b) shows the initial model employed in all FWI tests, which is kinematically inaccurate and a "generator" of cycle-skipped data. This model was constructed by applying a Gaussian smoothing operator with a standard deviation of 800 m. We carried out data inversions by considering the conventional FWI and the  $\kappa$-GSOT-FWI in the limit $\kappa \rightarrow 0$ and for $\kappa = 0.6$ as suggested by Ref. \cite{DASILVA2022127554}. Furthermore, in all numerical experiments, we employ 50 FWI iterations by using a conjugate gradient method. We also take into account two distinct scenarios for the type of noise in the dataset. In the first one, we consider a dataset contaminated by Gaussian noise with a signal-to-noise ratio (SNR) of 30. In the second scenario, we consider the same dataset as the first; however, we add a collection of erratic data with various amplitudes to simulate a non-Gaussian noise. In particular, we polluted $3\%$ of the seismic traces with spikes ranging from $5\,\zeta$ to $20\,\zeta$, in which $\zeta$ represents a standard normal random variable.

\begin{figure*}[!t]
  \centering
  \includegraphics[width=\textwidth]{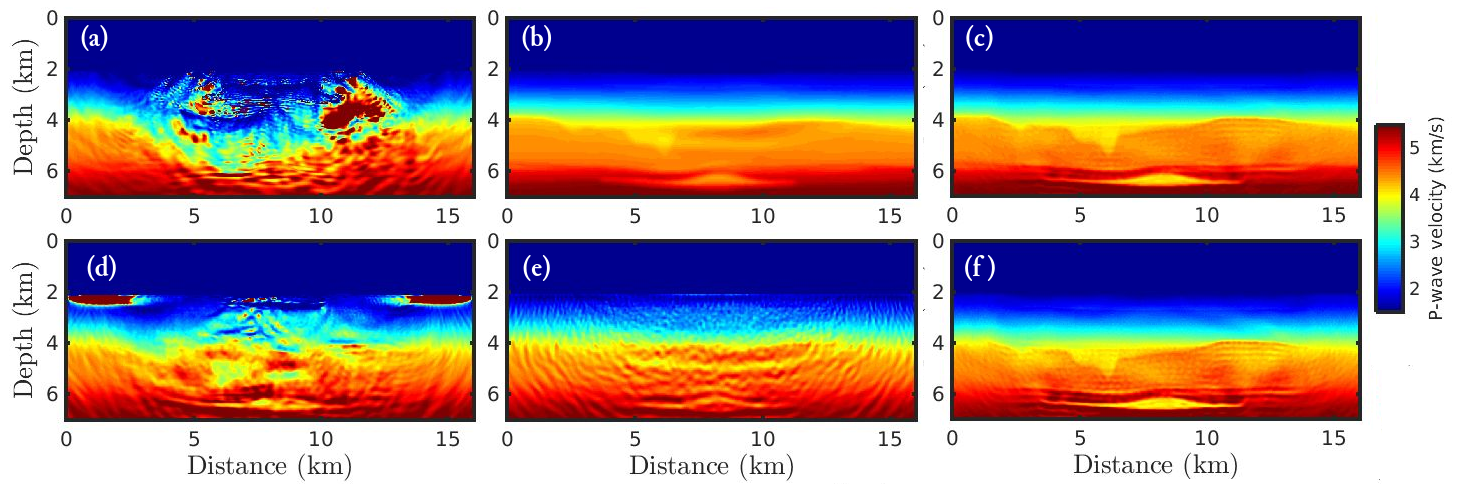}
  \caption{Recovered models for the Gaussian noise case by employing the (a) conventional FWI approach and the $\kappa$-GSOT-FWI framework with (b) $\kappa \rightarrow 0$, and (c) $\kappa = 0.6$. Recovered models for the non-Gaussian noise case by employing the (d) conventional FWI approach and the $\kappa$-GSOT-FWI framework with (e) $\kappa \rightarrow 0$, and (f) $\kappa = 0.6$.}
    \label{fig:models_resulting}
\end{figure*}

\begin{figure}[!htb]
  \centering  \includegraphics[width=\columnwidth]{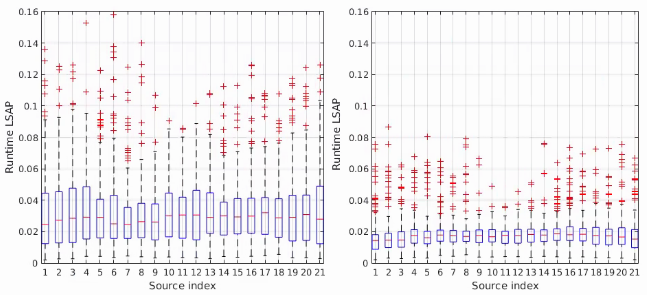}
  \caption{Runtime to resolve the LSAP. Top panel refers to case $\kappa \rightarrow 0$, while bottom panel refers to the $\kappa=0.6$ case.}
    \label{fig:runtime}
\end{figure}

Panels (a)-(c) of Fig. \ref{fig:models_resulting} show the recovered FWI models for the Gaussian noise case, while Panels (d)-(f) show the recovered FWI models for the non-Gaussian noise case. From a visual inspection, it is evident that the conventional FWI approach fails, as expected, to obtain an at least reasonable subsurface model when the initial model is kinematically inaccurate and/or when the data are contaminated by non-Gaussian noise, as depicted in Figs. \ref{fig:models_resulting}(a) and \ref{fig:models_resulting}(d). On the other hand, the $\kappa$-GSOT-FWI approach in case $\kappa \rightarrow 0$ reasonably recovers the velocities model when the data is contaminated with Gaussian noise, mitigating the cycle-skipping impacts, as depicted in Fig. \ref{fig:models_resulting}(b). However, the $\kappa$-GSOT-FWI with $\kappa \rightarrow 0$ would need several other FWI iterations to converge to a better model, spending more time and computational resources. In contrast, using the $\kappa$-Gaussian distribution in simultaneous with the OT metric quickly converges to an accurate final model, regardless of whether the noise is Gaussian, as depicted in Figs. \ref{fig:models_resulting}(c) and \ref{fig:models_resulting}(f). In fact, the classical approaches (conventional FWI and $\kappa$-GSOT-FWI with $\kappa \rightarrow 0$), completely fail to recover a good velocity model when non-Gaussian noise is considered, as depicted in Figs. \ref{fig:models_resulting}(a)-(b) and \ref{fig:models_resulting}(d)-(e).

Figure \ref{fig:runtime} shows the runtime for computing the combinatorial problem, the linear sum assignment problem (LSAP), for each seismic trace, in which each box plot represents a receiver-gather. Panel (a) refers to the classical approach, while Panel (b) refers to our proposed methodology. Analyzing the box plots, it is remarkable that the runtime for obtaining the LSAP solution for the $\kappa$-GSOT-FWI with $\kappa = 0.6$ is faster, saving computational resources and delivering better models.

\section{Conclusions}

We investigated in this work the portability of a new objective function in the FWI context based on the graph-space optimal transport framework and the $\kappa$-generalized Gaussian statistics. In this regard, we examined the robustness of the $\kappa$-GSOT-FWI in mitigating the impacts of cycle-skipping problems and non-Gaussian errors. The deepwater Brazilian pre-salt numerical example confirmed that the classical approach is sensitive to the quality of the initial model and non-Gaussian errors. Our findings also revealed that the proposed $\kappa$-GSOT-FWI with $\kappa = 0.6$ outperforms the classical case $\kappa \rightarrow 0$, mitigating the negative influence of cycle-skipping ambiguity and non-Gaussian noises. Our analysis also showed that the assumption that the errors follow a $\kappa$-Gaussian distribution with $\kappa = 0.6$ reduces the computational execution time of the computation of the transport plan, that is, of the LSAP solution. Although previous works have demonstrated that $\kappa = 0.6$ is interesting, an analytical investigation on the choice of the hyper-parameter $\kappa$  must be carried out.


\end{document}